\documentclass[preprint]{emulateapj}
\usepackage{amsmath,amssymb}
\usepackage{graphicx}
\usepackage{natbib}
\usepackage{color}
\usepackage[colorlinks,citecolor=blue,linkcolor=red]{hyperref}
\usepackage{threeparttable}
\usepackage{rotating}

\def\s{{\rm\thinspace s}}
\def\km{{\rm\thinspace km}}
\def\Mpc{{\rm\thinspace Mpc}}

\def\kmps{\hbox{$\km\s^{-1}\,$}}
\def\kmpspMpc{\hbox{$\kmps\Mpc^{-1}$}}

\begin{document}

\title{Discovery of a Mid-infrared Echo from the TDE candidate in the nucleus of ULIRG F01004-2237}

\author{Liming Dou\altaffilmark{1,2}, Tinggui Wang\altaffilmark{3}, 
Lin Yan\altaffilmark{4,5}, Ning Jiang\altaffilmark{3}, 
Chenwei Yang\altaffilmark{3}, Roc M. Cutri\altaffilmark{5}, Amy Mainzer\altaffilmark{6} and Bo Peng\altaffilmark{3}}

\altaffiltext{1}{Center for Astrophysics, Guangzhou University, Guangzhou 510006, China; doulm@gzhu.edu.cn}
\altaffiltext{2}{Astronomy Science and Technology Research Laboratory of Department of Education of 
	Guangdong Province, Guangzhou 510006, China}
\altaffiltext{3}{CAS Key Laboratory for Researches in Galaxies and Cosmology,
	University of Sciences and Technology of China, Hefei, Anhui 230026, China; twang@ustc.edu.cn}
\altaffiltext{4}{Caltech Optical Observatories, Cahill Center for Astronomy and Astrophysics, 
	California Institute of Technology, Pasadena, CA 91125, USA; }
\altaffiltext{5}{IPAC, Mail Code 100-22, California Institute of Technology, 1200 E. California Boulevard, Pasadena, 
	CA 91125, USA; lyan@ipac.caltech.edu}
\altaffiltext{6}{Jet Propulsion Laboratory, California Institute of Technology, Pasadena, CA 91109, USA}

\begin{abstract}

We present the mid-infrared (MIR) light curves (LCs) of a tidal disruption event (TDE) candidate
in the center of a nearby ultraluminous infrared galaxy (ULIRG) F01004-2237
using archival {\it WISE} and {\it NEOWISE} data 
from 2010 to 2016.
At the peak of the optical flare, F01004-2237 was IR quiescent. About three years later, 
its MIR fluxes have shown a steady increase, rising by 1.34 and 1.04 mag in $3.4$ and $4.6\mu$m
up to the end of 2016. The host-subtracted MIR peak luminosity is $2-3\times10^{44}$\,erg\,s$^{-1}$.
We interpret the MIR LCs as an infrared echo, {\it i.e.} dust reprocessed
emission of the optical flare. Fitting the MIR LCs using our dust model,
we infer a dust torus of the size of a few parsecs at some inclined angle.
The derived dust temperatures range from $590-850$\,K, and 
the warm dust mass is $\sim7\,M_{\odot}$. Such a large mass implies that the dust cannot be newly formed.
We also derive the UV luminosity of $4-11\times10^{44}$\,erg\,s$^{-1}$. 
The inferred total IR energy is $1-2\times10^{52}$\,erg, suggesting a large
dust covering factor. 
Finally, our dust model suggests that the long tail of the optical flare could be due to dust scattering.

\end{abstract}




\section{Introduction} \label{sec:intro}

When a star passes within a tidal radius ($r_{\rm t}\,=\,({M_{\rm BH}}/{M_{\rm *}})^{1/3}\,{r_*}$)
of a supermassive black hole (SMBH), 
it will be torn apart by tidal forces. 
About half of the stellar material is on unbound orbits and escapes, and the other half
is accreted by the central black hole, generating luminous flares in the soft X-ray, UV, 
and optical wavelengths, lasting a few months to years\,\citep{Rees1988,Rees1990,Hills1975,Lidskii1979}.
This is known as tidal disruption event (TDE). 
For a solar-type star and a non-rotating black hole $>$\,10$^8$\,$M_\odot$, 
the tidal radius will be inside the event horizon, then a TDE is not observable.
So most TDEs occur in lower-mass SMBHs ($<$\,10$^8$\,$M_\odot$). 
A dozen TDEs have been reported so far from transient surveys or serendipitous 
observations from X-ray to optical (see \citealp{Komossa2015} for a recent review).

TDEs serve as a powerful probe of the black hole accretion process itself. In addition,
they offer a unique opportunity to study interstellar medium within a few parsecs 
surrounding quiescent SMBHs. The UV and soft X-ray flares from TDEs can ionize gas medium, 
and the subsequent recombination produces spectral signatures. These spectral features are detected as 
transient extreme coronal lines, variable [O\,{\sc iii}] lines, 
or broad transient {He\,{\sc ii}} and H$\alpha$ 
lines \citep{Komossa2008,Wang2011,Wang2012,Gezari2012,Yang2013,Arcavi2014,Holoien2016}. 
When dust is present, UV/optical photons from TDEs are expected to be absorbed and re-emitted 
in the infrared. This dust emission is predicted to peak at 3-10$\mu$m with an infrared 
luminosity of 10$^{42-43}$\,erg\,s$^{-1}$, and last for a few years for a typical TDE \citep{Lu2016}. 
Such infrared emission has been detected for the first time recently in 
TDE candidates 
\citep{Dou2016,Jiang2016,van2016} using archival data from {\it Wide Field Infrared Survey Explorer} (WISE) 
and {\it Near Earth Object Wide-field Infrared Survey Explorer Reactivation} (NEOWISE-R) survey \citep{Wright2010,Mainzer2014}.

So far, most TDEs are found in normal post-starburst galaxies, whose star formation has been recently 
shut off \citep{French2016}. However, \cite{Tadhunter2017} recently 
reported the first case of a TDE candidate in a 
ULIRG F01004-2237 at $z=0.1178$. They found a luminous optical flare with 
a $L_{peak}\,\sim\,3\times10^{43}$\,erg\,s$^{-1}$ 
(peak date 2010 June 25) using the data from 
the Catalina Sky Survey (CSS, \citealp{Drake2009}). 
Following the optical flare, broad {He\,{\sc ii}} emission lines, were also detected in 
F01004-2237 \citep{Tadhunter2017}, which have been considered as the hallmarks of TDEs discovered 
from optical transient surveys \citep{Arcavi2014,Blagorodnova2017}. 
In addition, the estimated black hole mass is $\sim\,2.5\times10^7\,M_\odot$ \citep{Dasyra2006}. 
Based on these characteristics, 
this optical flare was classified as a TDE candidate (Tadhunter et al. 2017).

This TDE candidate is unusual.
First, the host galaxy has star formation rate $>\,100\,M_\odot\,yr^{-1}$ with massive young Wolf-Rayet stars 
in its compact nucleus \citep{Surace1998}.  
Both optical spectral line ratios and {\it Spitzer} MIR spectra suggest that 
it is a Seyfert 2 galaxy with a very dusty active galactic nucleus \citep[AGN;][]{Veilleux2009,Yuan2010}.
Second, the optical light curve (LC) 
has an unusually long decay time scale. Furthermore, three years after the peak, 
the LC did not fade back to its quiescent state. 
Instead, it levels off and is about 0.1 mag brighter.
These unusual features cast many questions as to the true nature of this event. 
Could the optical flare be simply AGN variability? Is it possible that broad {He\,{\sc ii}} 
lines could be the result of variable accretion rates?

In this Letter, we report the MIR flare detected in F01004-2237 three years after 
the optical peak date. 
The $3.4$ and $4.6\mu$m LCs are currently still rising. 

The Letter is organized as follows. The MIR LCs and results are described 
in Section\, 2. We discussed the results and concluded in Section\, 3 and 4. Throughout this Letter, 
we adopt a ${\Lambda}CDM$ cosmology with $\Omega_{M}$\,=\,0.3, $\Omega_{\Lambda}$\,=\,0.7, and a Hubble 
constant of $H_{0}$\,=\,70\,$\kmpspMpc$. With the redshift of F01004-2237 ($z=0.1178$), it results in 
a luminosity distance of 548.8 $Mpc$.

\section{Data Analysis and Result} \label{sec:datares}

\subsection{Light Curves}\label{sec:lc}

\begin{figure}
\figurenum{1}
       \centering
        \begin{minipage}{0.48\textwidth}
        \centerline{\includegraphics[width=1.0\textwidth]{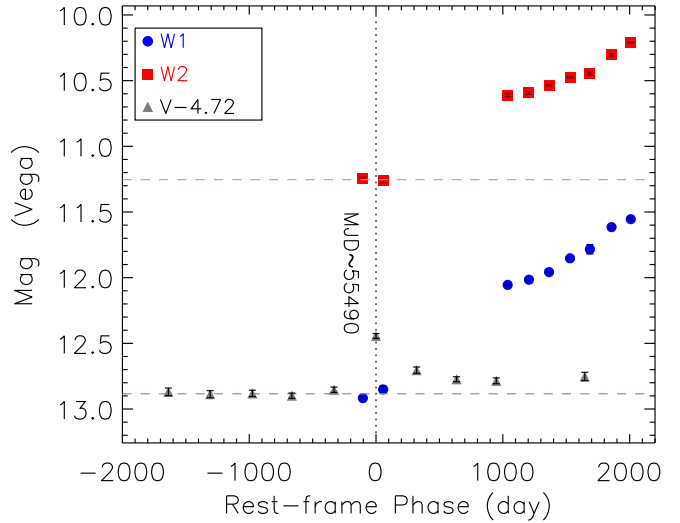}}
        \end{minipage}
        \caption{Light curves in V, W1, and W2 band of F01004-2237. The W1 and W2 band data are from 
	{\it WISE} and {\it NEOWISE-R}, the V band data monitored by CSS are from \cite{Tadhunter2017}. 
	The rest-frame phase is relative to the date of the optical flare peak detected by the 
	CSS (dotted line). The dashed lines are the average magnitudes in W1 and W2 bands of
        quiescent state. The V band light curve has been scaled down by 4.72\,mag for clarity. \label{fig:lcs}}
\end{figure}

The primary {\it WISE} survey ended on 2011 February 1, when
it was put into hibernation \citep{Wright2010}. 
The {\it NEOWISE-R} survey began on 2013 Dec 13 \citep{Mainzer2014}.
There was a $\sim3$ year gap between {\it WISE} and {\it NEOWISE-R}.
We extract a total of nine epochs of MIR photometry from the {\it WISE/NEOWISE} archive.
There are 10-20 exposures in each epoch. We use only the best-quality single-frame images 
by selecting only detections with data quality flag `qual$\_frame$'\,$>$\,0. 
This leaves 9-17 measurements for each epoch.
F01004-2237 is a point-like source in all of the {\it WISE} images.
Checking for potential contamination, we do not find any other sources within 
10\arcsec.
We do not find short-term variabilities within the individual epochs. 
We then average these fluxes to obtain a mean value at each epoch. 
We checked the MIR LCs of three other point sources close to our target, and no variability is detected in them.
The W1 and W2 band LCs are shown in Figure\,\ref{fig:lcs}. 
The V band LC from the CSS is also overplotted and scaled down by 4.72\,mag for clarity \citep{Tadhunter2017}.

There are no significant variations in the W1 and W2 bands between the first two epochs, $\sim120$ days before and  
$\sim60$ days after the peak of optical flare detected by the CSS. 
At the first epoch, the {\it WISE} 4-band photometry is $12.88\pm0.02$, $11.26\pm0.02$, $6.08\pm0.02$, 
and $3.10\pm0.02$\,mag at W1, W2, W3, and W4 bands, respectively, which is consistent with MIR fluxes 
derived from the broadband IR SED including the {\it Spitzer} IRS spectrum (5-35\,${\mu}$m) taken in 2004 
and the archival 2MASS JHK photometry \citep{Veilleux2009}.
This implies that the MIR emission is in a quiescent state during the first two epochs. 
The averaged brightness over these two epochs at $3.4$ and $4.6\mu$m is $12.88\pm0.04$ and $11.26\pm0.02$\,mag.

\begin{figure}
\figurenum{2}
       \centering
        \begin{minipage}{0.48\textwidth}
        \centerline{\includegraphics[width=1.0\textwidth]{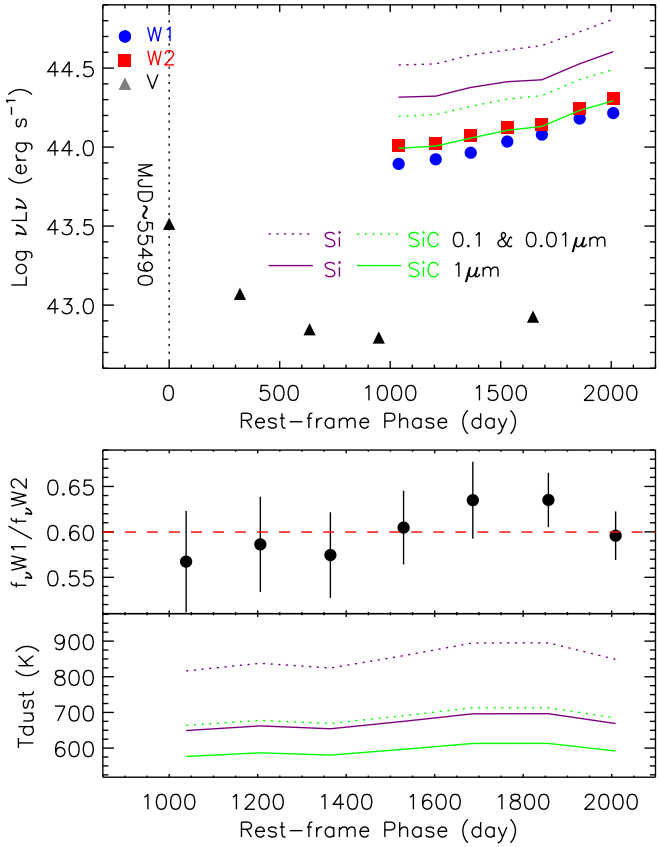}}
        \end{minipage}
        \caption{Upper panel: evolution of the host subtracted luminosity in V, W1, and W2 bands of F01004-2237. 
   	  The predicted 1-10${\mu}m$ luminosities for Si (purple) 
          and SiC (green) dust grains of varying sizes are also plotted.
          Middel panel: evolution of the ratio of host-subtracted flux between W1 and W2 bands. 
          The red dashed line is the mean level.  
          Lower panel: evolution of the dust temperature. 
          The uncertainty of the dust temperature is typically 30-40\,K.  
          \label{fig:excesslum}}
\end{figure}

When F01004-2237 was re-observed during the {\it NEOWISE-R} survey $\sim$3\,years after the 
optical flare peak, it was 
0.83 and 0.64 mag brighter in W1 and W2 bands, compared 
with the quiescent state. 
The source has been brightening steadily up to 2016 December 13 
(the last epoch in Figure \ref{fig:lcs}). Over the course of $\sim$6\,years after the optical flare, 
the W1 and W2 magnitudes have brightened by 1.35 and 1.06 mag, respectively. 
After subtracting the quiescent flux, 
we show the MIR LCs in 
the upper panel of Figure \ref{fig:excesslum}. For
comparison, we also show the LC of V band after subtraction of the constant component
defined by pre-flare luminosity. 

The host-subtracted, maximum MIR luminosity with the available data 
is $10^{44.2}$ and $10^{44.3}$\,erg\,s$^{-1}$ at $3.4$ and $4.6\mu$m, respectively.
This is a factor of 6 higher than the peak luminosity of the optical flare (host-subtracted).
The strong silicate absorption in the pre-flare {\it Spitzer} MIR spectrum 
indicates this is a heavily dust obscured Seyfert 2 galaxy \citep{Veilleux2009,Yuan2010}.
Therefore, the optical fluxes likely suffer from strong dust extinction.

\subsection{Dust Temperature and MIR Luminosity} \label{sec:temlum}

Because infrared fluxes are available only in the W1 and W2 bands 
during the {\it {NEOWISE-R}} survey, we rely on models to estimate dust temperature.
For simplicity, we consider a single-temperature dust consisting of either astronomical silicate (Si) 
or silicon carbide (SiC) with three different grain sizes, 0.01, 0.1, and 1\,$\mu$m 
({$f_{\nu} = Q_{abs}(\nu)\,B\,({\nu},\,T_{d})$). 
The absorption efficiency  $Q_{abs}(\nu)$ for the dust is taken from \cite{Laor1993}.

Applying these models to the host-subtracted W1 and W2 light curves, we estimate dust temperature 
and integral luminosity in 1-10\,$\mu$m at each epoch (see also Figure~\ref{fig:excesslum}). 
The estimated dust temperatures appear to be almost constant as a function of 
time for a single grain size. This is due to the fact that the host-subtracted W1 
and W2 flux ratios are $\sim$0.6, roughly constant over time.
The derived T$_d$ ranges from 590 to 850\,K with the assumed grain size of 0.01-1$\mu$m, respectively.  
Finally, the luminosities derived from dust emission are 
in a narrow range of $(2-5)\,\times10^{44}$\,erg\,s$^{-1}$, at the last epoch. 

\section{Discussion}

\subsection{Dust Echo of a TDE?}

Extragalactic infrared variable sources can be categorized into two groups: non-thermal 
emission from relativistic jets such as blazars,  and thermal dust emission as in radio quiet 
AGNs and TDEs (Wang et al. 2017).
Less than 10\% of the known TDEs show jets, and 
the emission from a beamed jet often shows dramatic variability on time scale of hours to days\,\citep{Jiang2012,Mangano2016}
However, F01004-2237 is radio quiet\,\citep{Rodriguez2013}, during 3-6 years after the optical peak,
its flux ratio between W1 and W2 band remains constant, 
and there are no significant short-term MIR variability.
Thus,  the observed infrared emission is unlikely to be non-thermal. 
For the remaining discussion, we consider only the thermal emission scenario.
A UV flare, regardless of its origin as a changing-look 
AGN or a TDE, will cause a response in MIR emission if there is dust around the black hole at parsec distance scale. 
This is a so-called infrared echo. However, the MIR LC of F01004-2237 looks quite 
different from those of other TDEs/TDE candidates \citep{Dou2016,Jiang2016,van2016}. 
The question here is what conditions would produce an LC like this.

First, there is no significant MIR excess above the pre-flare level at 60 days after 
the optical flare. In other words, the UV/optical photons from the flare had not been 
reprocessed by dust in 60 days. 
This implies that there must be very little dust reprocessing the UV flare along the 
line of sight within a radius of 30 lt-days from the optical flare source 
(Figure \ref{fig:cartoon})\,\footnote{Strict speaking, there is little dust within the 60 day iso-lag surface. An iso-lag 
surface is a paraboloid with the UV source being its focus and the pericenter distance being $c\tau/2$.}. 
Lack of dust within the radius of 30 lt-days at the peak of optical flare is expected because the 
dust sublimation radius is $>30$ lt-days when the peak luminosity of the optical flare 
exceeds\footnote{The dust sublimation radius is $R_{sub}\,=\,
0.47\,(6{\nu}L_{\nu}(V)\,/\,(10^{46}\,$erg\,s$^{-1}))^{1/2}$\,$pc$ \citep{Kishimoto2007,Koshida2014}.} $10^{43}$\,erg\,s$^{-1}$. 

Second, our modeling of the data shows that the dust temperature remains roughly constant, 
suggesting that UV radiation intensity from the central source is also roughly the same at all epochs. 
Since the optical flare lasts a much shorter time scale than that of MIR LC, 
we can simplify and consider the optical flare as a pulse. 
Therefore, we infer that the dust distribution must be in a ring shape with equal 
distances to the central heating source. As shown in Figure~\ref{fig:cartoon}, 
when the UV photons from the central source travel further out, 
they reach more and more surface areas of the dust torus, heating up more dust and producing 
a rising MIR LCs. If this dusty ring is viewed at an intermediate angle,
it can explain the lack of delay at less than 60 days and at the same time 
reproduce the continuous rise of the MIR flux. This was demonstrated by the similar double-horn 
transfer function for an inclined ring calculated for broad-line reverberation mapping 
study \citep{Peterson2004}.
These horns are separated by $2r\,\cos\theta/c$, where $r$ is the radius of the ring and 
$\theta$ is the inclination angle of the ring.
Unfortunately, we do not have the early stage MIR LC, but one prediction of the model 
is that MIR flux will turn over sharply after a certain time (see the right panel of Figure~\ref{fig:cartoon}). 
This can be tested with future MIR 
observations. The radius of the ring is at least 3 lt-yr, but not much larger, in order to match 
the LC.   

\begin{figure*}
\figurenum{3}
       \centering
        \begin{minipage}{0.45\textwidth}
        \centerline{\includegraphics[width=1.0\textwidth]{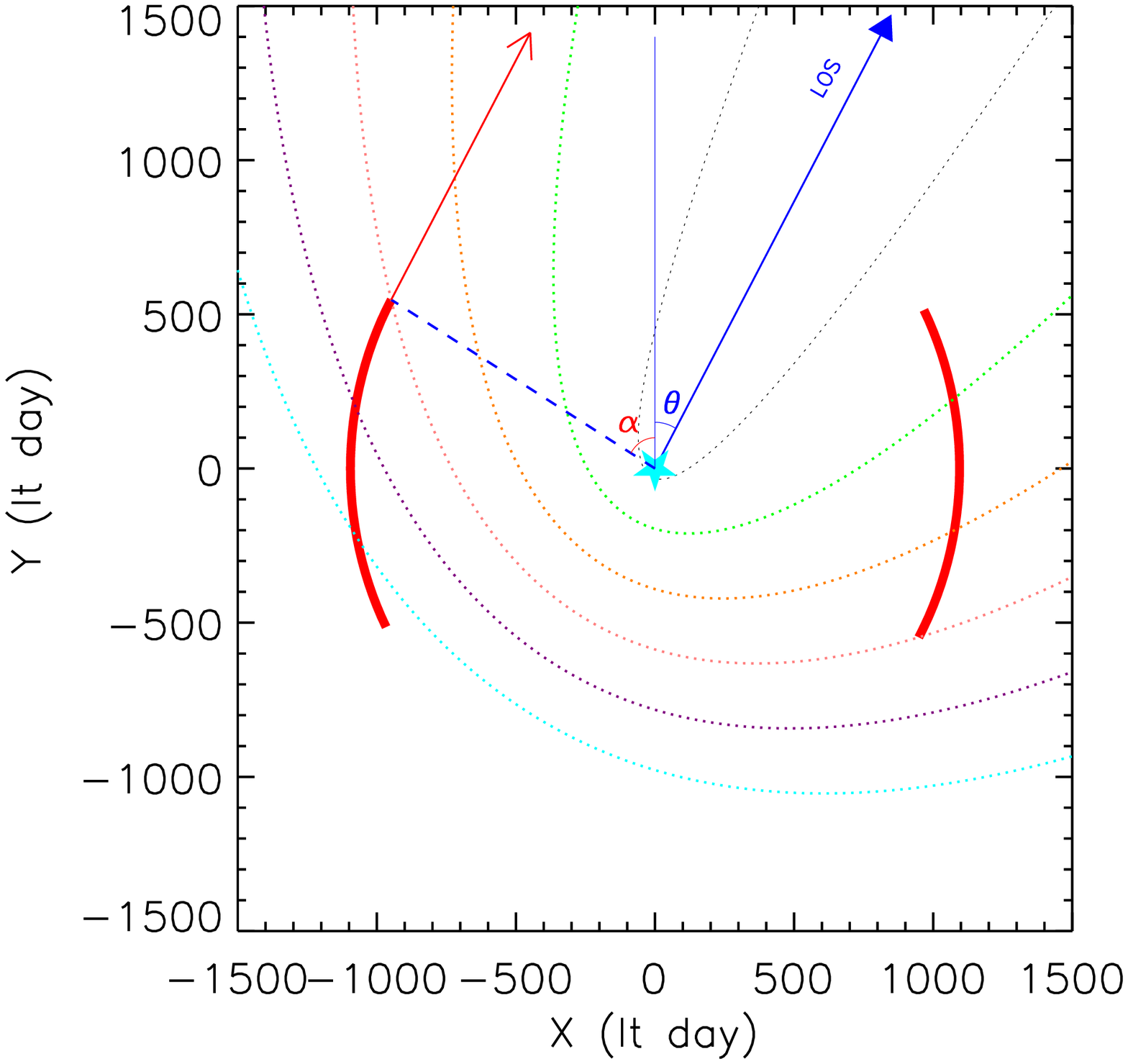}}
        \end{minipage}
        \begin{minipage}{0.48\textwidth}
        \centerline{\includegraphics[width=1.0\textwidth]{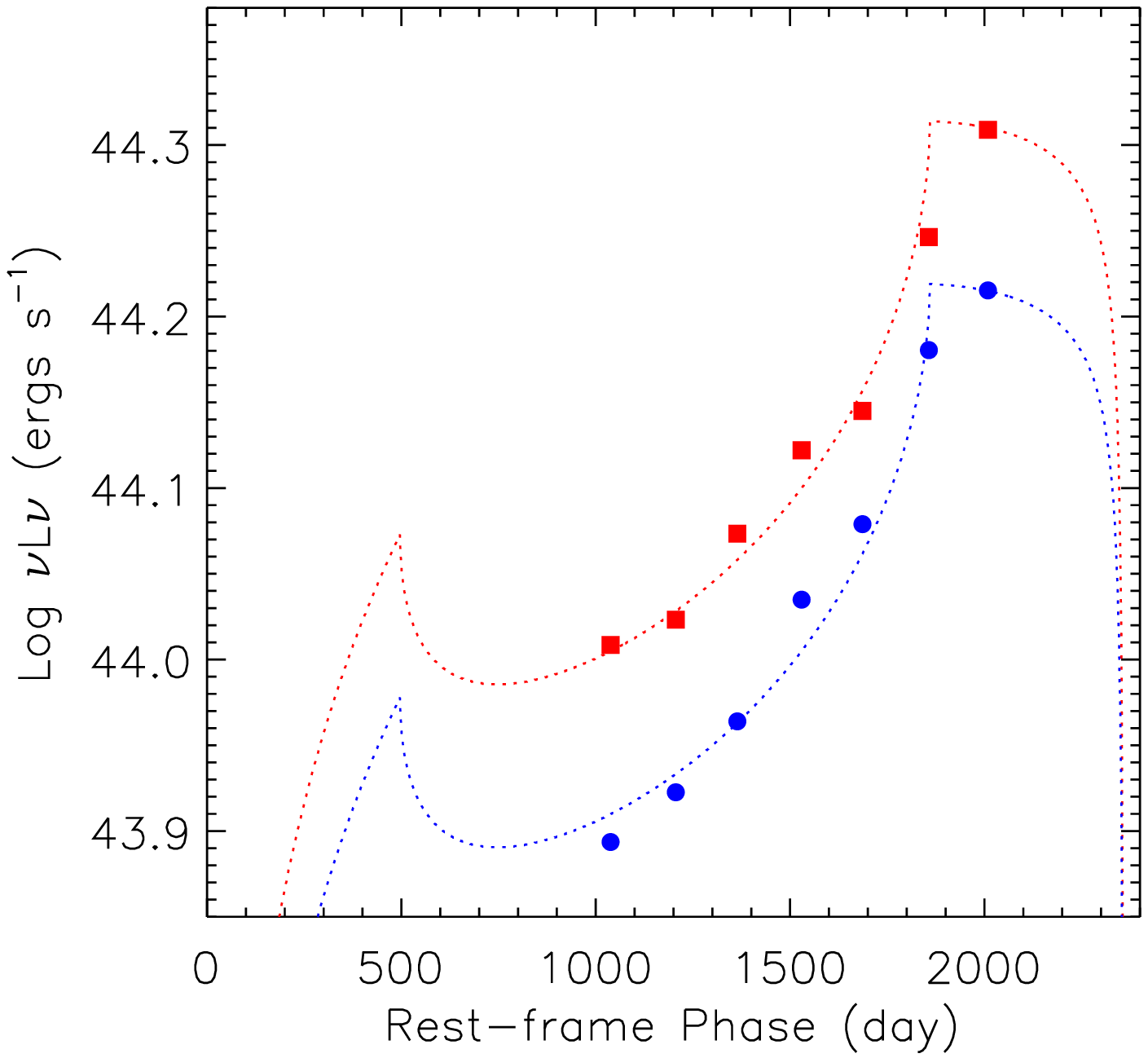}}
        \end{minipage}
        \caption{Left panel: schematic of a convex dusty ring (red bows) that absorbs UV photons 
        and re-emits in the infrared band and, simultaneously, scatter UV and optical photons into 
        our line of sight. The dashed lines illustrate the iso-delay surface at lags of 60 days, 
        1, 2, 3, 4, and 5 years.
        Right panel: example of theoretical light curves for the convex dusty ring model. 
        We assume that infrared emission follows $I\propto \cos \Theta$, i.e., in optically thick case. 
        The ring parameters are: $\theta$\,=\,45$\arcdeg$, $\alpha$\,=\,70$\arcdeg$, and r\,=\,1.13\,$pc$.
        The red (blue) dotted line is for W1 (W2) band.
         \label{fig:cartoon}}
\end{figure*}

Another interesting inference from the dust reprocessing scenario is the prediction of 
a dust echo signature (reflection component) in both the optical and UV bands. If dust 
composition and grain size are uniform over the region, we expect that the scattered 
optical flux is roughly proportional to that of reprocessed infrared light. The fraction of 
scattered light depends on the albedo of the dust, which is a strong function of grain 
size. 
Grains of size $>0.1\mu$m will give strong reflection light up to 1 $\mu$m. With a 
reasonable albedo of dust in V band \citep[$\sim$0.5;][]{Laor1993} and optical to 
UV light ratio of incident continuum ($\sim$0.2), scattered optical flux would be 
$\sim$10\% of the 
infrared flux. This is sufficient to explain the puzzling up-turn in the optical LC 
several years after the optical flare.
 
\subsection{The UV Luminosity and Lower Limits to the Total Energy Release during the Flare} \label{sec:uv}

Dust grains can be considered as bolometers due to large absorption cross-section in UV. 
Their temperature is determined by the incident UV flux. Its UV luminosity can be inferred from 
dust temperature and distance from the UV source. As discussed in the last section, 
the heated dust is likely in a convex surface with a distance to the UV source of at 
least 3 lt-yr. Using the formulae (1-3) in \cite{Dou2016} and adopting $Q_{abs}$ from \cite{Laor1993}, 
one can estimate the UV luminosity.  

For Si grains, the UV luminosity calculated from the best-fit convex dusty ring model is 
(0.06, 0.6 and 4.6)$\times10^{44}$\,erg\,s$^{-1}$ for grain size of 0.01, 0.1, and 1\,$\mu$m, respectively.
The value will be 1.4-2.2 times larger for SiC grains. The IR spectrum observed at any time is actually 
the sum of the emission components from where the dust ring intersects different iso-delay surfaces.
Different areas are illuminated with different levels of UV luminosity and therefore have a range of 
equilibrium temperatures. Consequently, the single-temperature model described above underestimates 
the peak UV luminosity by a factor of 2-3. 
MIR spectra of the flare would allow us to constrain the dust properties. 
This should lead to a more better estimate of the UV luminosity. 
Then, we will consider another way to estimate the integrated UV luminosity.

We calculate the total energy emitted in the infrared flare by integrating the 
host-subtracted LC. During the {\it NEOWISE-R} survey, F01004-2237 has emitted 
a total IR energy of about $1.3-2.0\,\times\,10^{52}$\,erg.
The dust covering factor is about 0.34, inferred from the opening angle ($\alpha$\,=\,70$\arcdeg$) of the best-fit
convex dusty ring model. If the total IR energy comes from the heated by optical/UV photons that released
during the optical/UV flare, the total optical/UV energy should be at least $4\,\times\,10^{52}$\,erg. 
This is about three times the energy released by a tidal 
disruption of a $0.3\,M_{\odot}$ star proposed by \cite{Tadhunter2017}, assuming half of the 
debris is accreted with a typical radiation efficiency of $\eta\,=\,0.1$. 
Given the fact that the early rising part of the LC was not observed, 
and the source is still brightening in MIR, the above energy is a lower limit.
This suggests that the disrupted star is likely to be more massive or that the radiation efficiency is higher than 0.1.
If the time scale of the optical/UV flare is one year \citep{Tadhunter2017}, the integrated UV luminosity 
should be at least $1.2\,\times\,10^{45}$\,erg\,s$^{-1}$. 
Roughly, this value is consistent with the one estimated from the dust temperature and 
dust ring model for the largest grains considered.


We estimate the heated dust mass at each epoch using the formula\,(8-9) in \cite{Dou2016}, 
adopting $\rho =2.5$\,g\,$cm^{-3}$ for Si. The heated dust mass is $\sim$5\,$M_\odot$ at the last epoch.
In our model, the dust was emitting MIR photons for the duration of the 
optical flare and then cools down. 
So the lower limit to the total amount of dust is $\sim7\,M_{\odot}$, 
if it is assumed the optical flare time scale is one year.
Such a large dust mass within a parsec radius of the SMBH implies that these dust grains cannot be
newly formed. They are likely already there from the dusty torus of the Seyfert 2 nucleus.

\section{Conclusions}

We report the discovery of a strong infrared echo of the optical flare from the TDE candidate 
in ULIRG F01004-2237 using archival {\it WISE} and {\it NEOWISE} data. 
The object is quiescent in the MIR during the peak of the optical flare in 2010. 
However, three years after the optical peak, we find a steady rise in MIR emission, 
with fluxes increasing by a factor of 3.4 and 2.6 at $3.4\mu$m and $4.6\mu$m, respectively.
The peak MIR luminosities are $1.6\,\times\,10^{44}$ and $2.0\times10^{44}$\,erg\,s$^{-1}$ in the W1 and W2 bands, 
respectively, which are about 0.7-0.8 dex higher than the observed peak flare luminosity in the V band. Despite the 
infrared luminosity continuously rising by a factor of 2 in the past three years, 
our model suggests that  
the temperature remained roughly constant at 590-850\,K. We estimate MIR 
energy radiated between 2013 and 2016 to be $(1.3-2.0)\,\times\,10^{52}$ erg. 
We interpret the variable infrared emission in the context of thermal dust emission in response to 
the UV/optical flare, presumably caused by a TDE. A thick dusty ring with a radius of about 
1 pc, and mass $>7\,M_\sun$, and grain size larger than 0.1\,${\mu}$m
around the black hole at an intermediate inclination angle may be able to reproduce all 
characteristics in the observed MIR LC.  
The model predicts a sharp decrease of the MIR flux not 
too far in the future, which can be verified by future photometric monitoring, and an optical component 
due to dust scattering of the primary UV/optical flare at a level 10\% of the infrared luminosity, 
which may explain the very flat optical LC at the very late stage, and can be tested with 
polarization observations.

We are grateful to the anonymous referee for comments that have improved the quality of this Letter.
We thank Wenbin Lu for helpful discussions and comments.
This research is supported by the National Basic Research Program of China (grant No.\, 2015CB857005), 
NSFC (NSFC-11233002, NSFC-11421303, NSFC-11603021), Joint Research Fund in Astronomy (U1431229, U1531245) under 
cooperative agreement between the NSFC and the CAS and the Fundamental Research Funds for the Central
Universities.
This research makes use of data products from the {\it Wide-field Infrared Survey Explorer (WISE)} and 
the {\it Near-Earth Object Wide-field Infrared Survey Explorer (NEOWISE)}. 
{\it WISE} is a joint project of the University of California, Los Angeles, 
and the Jet Propulsion Laboratory/California Institute of Technology; 
{\it NEOWISE} is a project of the Jet Propulsion Laboratory/California Institute of Technology. 
{\it WISE} and {\it NEOWISE} are funded by the National Aeronautics and Space Administration.


\end{document}